\documentclass{pasj00}

\begin{document}
\SetRunningHead{K. Otsuji et al.}{Magnetic Emergence Observed with Hinode/SOT}
\Received{2007/05/31}
\Accepted{2007/09/18}

\title{Small Scale Magnetic Flux Emergence Observed with Hinode/Solar Optical Telescope}

\author{
	Kenichi~\textsc{Otsuji},\altaffilmark{1,2}
	Kazunari~\textsc{Shibata},\altaffilmark{2}
	Reizaburo~\textsc{Kitai},\altaffilmark{2}
	Satoru~\textsc{Ueno},\altaffilmark{2}
	Shin'ichi~\textsc{Nagata},\altaffilmark{2}
	Takuma~\textsc{Matsumoto},\altaffilmark{1,2}
	Tahei~\textsc{Nakamura},\altaffilmark{1,2}
	Hiroko~\textsc{Watanabe},\altaffilmark{1,2}
	Saku~\textsc{Tsuneta},\altaffilmark{3}
	Yoshinori~\textsc{Suematsu},\altaffilmark{3}
	Kiyoshi~\textsc{Ichimoto},\altaffilmark{3}
	Toshifumi~\textsc{Shimizu},\altaffilmark{4}
	Yukio~\textsc{Katsukawa},\altaffilmark{3}
	Theodore~D.~\textsc{Tarbell},\altaffilmark{5}
	Bruce~W.~\textsc{Lites},\altaffilmark{6}
	Richard~A.~\textsc{Shine},\altaffilmark{5}
	and Alan~M.~\textsc{Title}\altaffilmark{5}
}
\altaffiltext{1}{Department of Astronomy, Kyoto University, Kitashirakawa-Oiwake-cho, Sakyo-ku, Kyoto 606-8502}
\email{otsuji@kwasan.kyoto-u.ac.jp}
\altaffiltext{2}{Kwasan and Hida Observatories, Kyoto University, Yamashina-ku, Kyoto 607-8471}
\altaffiltext{3}{National Astronomical Observatory, Mitaka, Tokyo 181-8588}
\altaffiltext{4}{Institute of Space and Astronautical Science, Japan Aerospace Exploration Agency, Sagamihara, Kanagawa 229-8510}
\altaffiltext{5}{Lockheed Martin Advanced Technology Center, Palo Alto, CA 94304, USA}
\altaffiltext{6}{High Altitude Observatory, 3450 Mitchell Lane, Boulder, 80301 Colorado USA}


\maketitle

\begin{abstract}
We observed small scale magnetic flux emergence in a sunspot moat region by the Solar Optical Telescope (SOT) aboard the Hinode satellite.
We analyzed filtergram images observed in the wavelengths of Fe 6302 {\AA}, G-band and Ca\emissiontype{II} H.
In Stokes I images of Fe 6302 {\AA}, emerging magnetic flux were recognized as dark lanes.
In G-band, they showed their shapes almost the same as in Stokes I images.
These magnetic flux appeared as dark filaments in Ca\emissiontype{II} H images.
Stokes V images of Fe 6302 {\AA} showed pairs of opposite polarities at footpoints of each filament.
These magnetic concentrations are identified to correspond to bright points in G-band/Ca\emissiontype{II} H images.
From the analysis of time-sliced diagrams, we derived following properties of emerging flux,
which are consistent with the previous works.
(1) Two footpoints separate each other at a speed of 4.2 km s$^{-1}$ during the initial phase of evolution and decreases to about 1 km s$^{-1}$ in 10 minutes later.
(2) Ca\emissiontype{II} H filaments appear almost simultaneously with the formation of dark lanes in Stokes I in the observational cadence of 2 minutes.
(3) The lifetime of the dark lanes in Stokes I and G-band is 8 minutes, while that of Ca filament is 12 minutes.

An interesting phenomena was observed that an emerging flux tube expands laterally in the photosphere with a speed of 3.8 km s$^{-1}$.
Discussion on the horizontal expansion of flux tube will be given with refernce to previous simulation studies. 
\end{abstract}

\section{Introduction}
Emerging flux region (EFR) is a young active region (AR) where magnetic flux loops emerge from underneath photosphere (\cite{Bru69}, \cite{Zir72}).
Emerging flux often causes reconnection with pre-existing coronal magnetic field and are responsible to flares (\cite{Hey77}, \cite{Shi92a}, \cite{Shi92b}, \cite{Shi99}, \cite{Shi02}).
\citet{Str96} observed the separation speed of both polarities of emerging flux to be less than 1km s$^{-1}$ by using MDI magnetogram.
They show that dark lanes are often observed in the photosphere of EFRs.
\citet{Har73} reported that separation speeds of loop footpoints are around 4 km s$^{-1}$ in the initial phase and reduced to 1 km s$^{-1}$ in the later phase of emergence.
The morphological features of emerging flux observed in H$\alpha$ are dark loops and bright points near their footpoints.
A cluster of dark loops is named as an "arch filament system" (AFS) by \citet{Bru67}.
A single arch filament is thought to be a trace of a magnetic flux tube.
The lifetime of an arch filament is 10-30 minutes and the rise velocity is 10-15 km s$^{-1}$ \citep{Cho88}.
Evolutions of emerging flux from the chromosphere to the corona were investigated by \citet{Yos99} and \citet{Shi02}.
They observed that emerging flux appeared in soft X-ray images about several minutes after the appearance of new arch filament in H$\alpha$ images.

\citet{Shi89} performed two dimensional magnetohydrodynamic (MHD) simulations of emerging flux for the first time,
which reproduced well various dynamical features of emerging flux,
such as the rise motion of an arch filament and donwflow along it.
Subsequently, three dimensional simulations of emerging flux have been carried by \citet{Mat92} and others (\cite{Mat93}, \cite{Fan01}, \cite{Mag01}, \cite{Noz05}).
These three-dimentional simulations showed that when there is no strong magnetic twist,
the emerging flux undergoes strong horizontal expansion in a direction perpendicular to emerging flux filament just after its emergence into the photosphere,
which are remained to be confirmed by actual observations.

Recently \citet{Par04} proposed a new way of emergence from the flare genesis experiment observation
that emerging flux rises in a form of undulatory loops in the photosphere, which must be confirmed observationally by many samples of EFRs. 
In this paper, we report the result of our analysis of an EFR observed by the Solar Optical Telescope (SOT) on board Hinode.
Thanks to multi-wavelength observation from space (\cite{Kos07}, \cite{Tsu07}, \cite{Sue07}, \cite{Ich07}, \cite{Shi07}), we could follow the temporal evolution of emerging loops in the photosphere and the chromosphere.
Our analysis is concentrated on (1) morphological evolution of an EFR in and below the chromospheric layers and (2) basic morphological characters of individual emerging loops.

\section{Observation and data reduction}
\subsection{Observation}
\begin{figure}
\begin{center}
\FigureFile(80mm,80mm){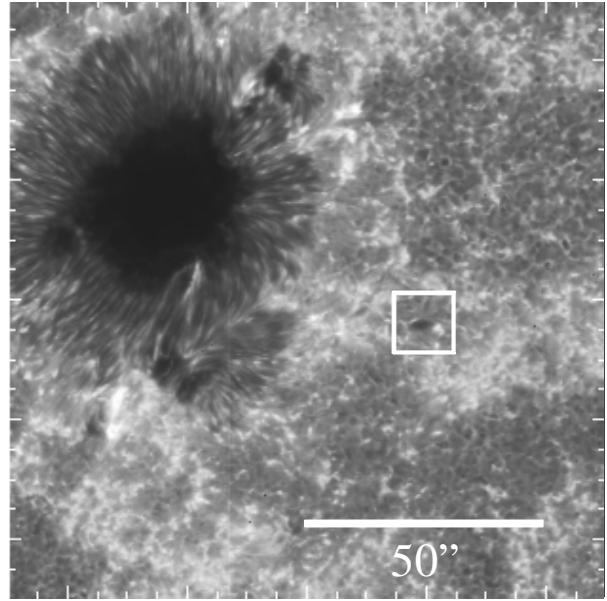}
\end{center}
\caption{
Ca\emissiontype{II} H line image of NOAA 10930 obtained by Hinode/SOT at 18:01:52 UT on December 19, 2006.
An EFR indicated by a white square was studied.
}
\label{fig1}
\end{figure}

Fig. \ref{fig1} shows an EFR appeared in the moat region of NOAA 10930 near the disk center (E\timeform{08D}S\timeform{05D}) on December 19, 2006 with Hinode/SOT.
The observation was done from 17:00 to 19:00 UT.
In this period, many small scale EFRs were observed at the south-west part of the region.

Multi-wavelength imaging observation was done in three wavelengths, i.e.,
(1) Fe 6302 {\AA} Stokes I and V,
(2) G-band 4305 {\AA}, and
(3) Ca\emissiontype{II} H 3968.5 {\AA}.
The observational cadence was 2 minutes.
The images in (1) were taken through the narrow-band filter imager (NFI) with the pass-band of 90{m\AA},
while the images in (2) and (3) through the broad-band filter imager (BFI) with the pass-band of 0.8{\AA} and 0.3{\AA}, respectively.
In our observation, spatial resolutions are 0.16"/pixel (NFI) and 0.10"/pixel (BFI).

\subsection{Data reduction}
\begin{figure}
\begin{center}
\FigureFile(80mm,40mm){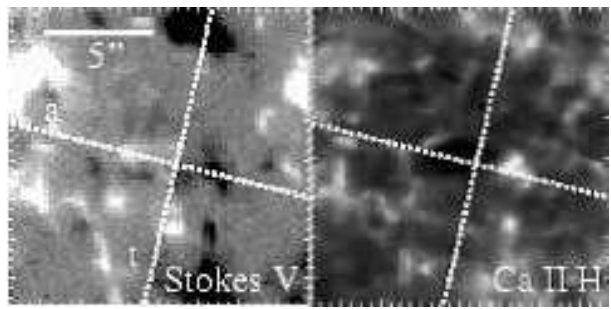}
\end{center}
\caption{
The emerging flux observed in Fe 6302 {\AA} Stokes V (left panel) and Ca\emissiontype{II} H (right panel).
Two dashed lines are drawn to represent the position of two slits for time-sliced diagrams.
Slit a is drawn on the two footpoints of the emerging flux (along-filament).
Slit t is perpendicular to the slit a and put on the midpoint of the two footpoints (trans-filament). 
}\label{fig2}
\end{figure}

Co-alignment between NFI and BFI images were done with reference to common granulation patterns.
To derive basic morphological properties of emerging loops,
we made two kind of time-sliced diagrams similar to \citet{Shi02}.
One is "along-filament" and the other is "trans-filament" time-sliced diagram,
as is described in Fig. \ref{fig2}.

\begin{figure}
\begin{center}
\FigureFile(80mm,120mm){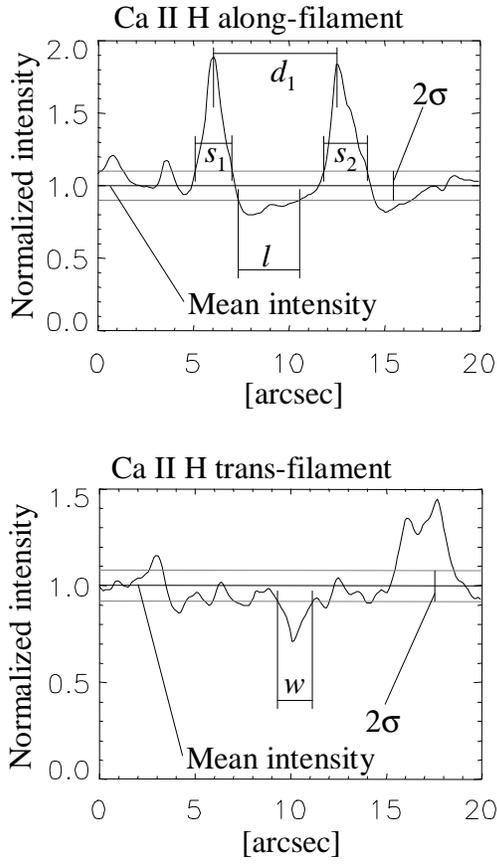}
\end{center}
\caption{Definition of linear sizes of Ca\emissiontype{II} H features.
The intensity distribution is normalized by the mean intensity of the neighboring quiet region.
$\sigma$ is the standard deviation of intensity fluctuation in the neighboring quiet region.
$d_1$: the distance between the two intensity peaks;
$s_1$ and $s_2$: the linear size of two bright points;
$l$: the length of the filament;
and $w$: the width of dark lane/filament.
The same definition is applied to Stokes I and G-band images.
}\label{fig3}
\end{figure}

\begin{figure}
\begin{center}
\FigureFile(80mm,60mm){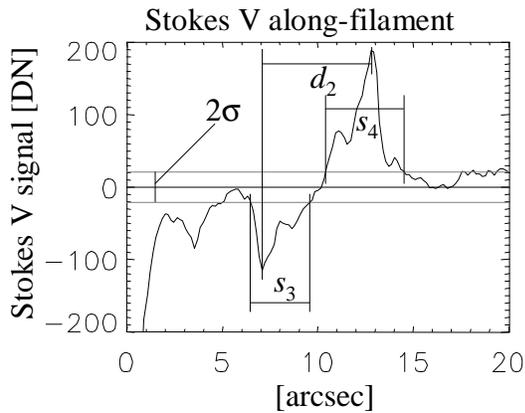}
\end{center}
\caption{Stokes V distribution on the along-filament slit.
The fluctuation of Stokes V signal on the neighborhood quiet region are expressed by $\sigma$.
In the panel, $d_2$: the distance between two opposite polarities of magnetic flux;
and $s_3$ and $s_4$: the linear size of footpoints of emerging flux. 
}\label{fig4}
\end{figure}

Fig. \ref{fig3} and \ref{fig4} shows the ways how to measure the various size of an emerging loop.
In this study, we analyzed the evolutions of
(1) the linear size of the footpoints of emerging loop,
(2) the distance between the footpoints,
and (3) the length and width of dark lane or filament.

\section{Morphological characteristics of the EFR}
\subsection{Before the emergence of the flux tube}

\begin{figure}
\begin{center}
\FigureFile(67mm,150mm){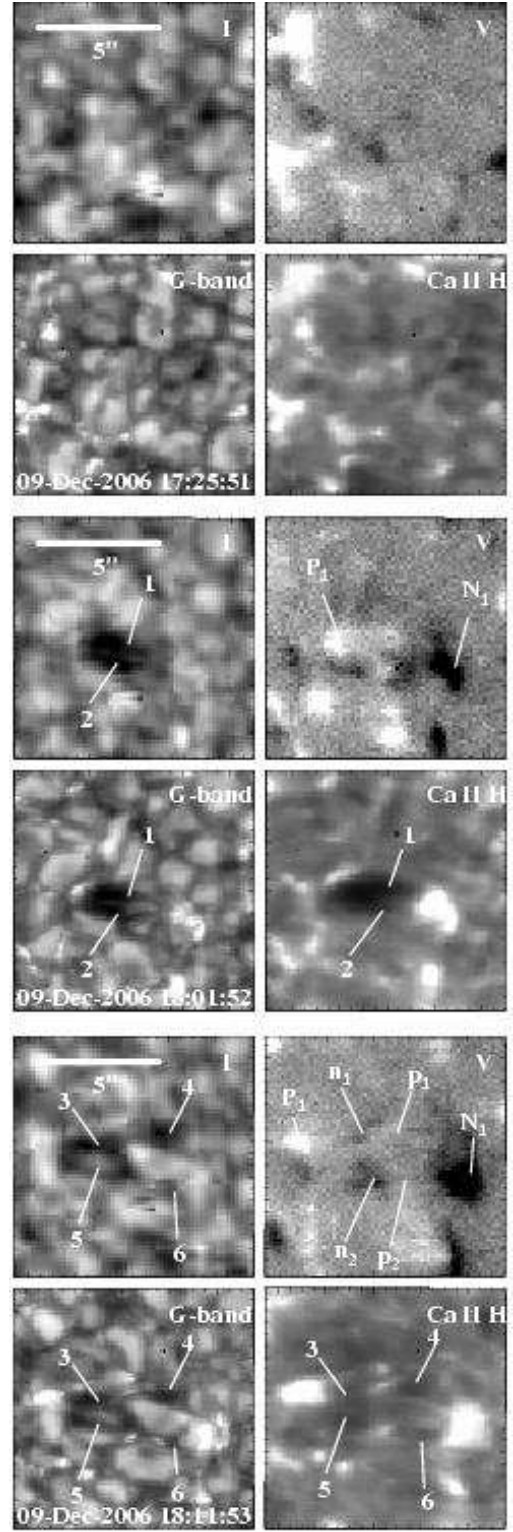}
\end{center}
\caption{
The flux tube at three stages of its emergence.
(a) before the emergence.
(b) the initial stage of the emergence.
(c) the later stage of the emergence.
In each 4 images, upper left, upper right, lower left and lower right show Fe 6302 {\AA} Stokes I, V,
G-band and Ca\emissiontype{II} images of the studied region with a common scale of magnification.
In the images, the loops of the EFR and magnetic polarities at the loop's footpoints are indicated by numbers and alphabets.
}\label{fig5}
\end{figure}

Fig. \ref{fig5} shows images of the flux tube at three stages of its emergence.
The upper 4 images are taken at 17:25:51 UT, when the flux emergence had not occurred yet.
Before the emerging flux's appearance, there was no particular feature in each 4 images.

\subsection{The initial stage of the emergence of the flux tube}
The middle 4 images of Fig. \ref{fig5} are taken at 18:01:52 UT,
when about 8 minutes passed from the beginning of the flux emergence.
In the image of Stokes I, there seems to be two dark lanes (1 and 2).
The widths of the dark lanes are about 1.5" and lengths of them are 4". 
There is no prominent brightening at the footpoints of the emerging flux.
In the image of Stokes V, we can see a pair of opposite polarities (P$_1$ and N$_1$) at the middle of the image.
They are the footpoints of the emerging flux.
The distance between two footpoints is about 5".
In G-band, we can identify two loops or filaments more clearly than in Stokes I.
The widths and the lengths of the arch filaments observed in G-band are almost the same as those in Stokes I.
At the both ends of arch filaments, there are tiny bright points in G-band.
The size of the bright points is about 0.5"$\times$0.5".
In the image of Ca\emissiontype{II} H, the arch filaments appear darker and wider.
Arch filament 1 and 2 almost merge to each other.
The width of individual arch filament is about 1" and the whole width of arch filaments is about 2".
The lengths of the arch filaments are about 4" and the same as in the other images.

\subsection{The later stage of the emergence of the flux tube}
The bottom 4 images in Fig. \ref{fig5} show the morphology of flux tube at its later phase of emergence.
The observation time is 18:11:53 UT, so 18 minutes after the emergence.
In the image of Stokes I we can see four dark lanes (3, 4, 5 and 6),
while there are no longer any dark lanes which connect directly between the original footpoints.
The widths of these dark lanes are about 0.7" and the length of each dark lane is about 2"-3".
In the image of Stokes V, the distance between two opposite polarities is 7" and larger than 10 minutes ago.
Note that in the middle region between the originally emerged footpoints there are several small magnetic concentrations (p$_1$, n$_1$, p$_2$ and n$_2$).
Same as in Stokes I image, there are four arch filaments in G-band image.
The values of the width and length of arch filaments are also same as those of Stokes I.
The size of the bright points at the footpoints is about 1"$\times$1".
In Ca\emissiontype{II} H image we can see four arch filaments, same as the other images.
The width of each arch filament is about 0.5", and the length of them is 2"-3".
At the middle of two main footpoints it forms new footpoints of the arch filaments.
While the size of the bright points at the main footpoints is 1.5"$\times$1.5",
those of the newly formed footpoints at the middle of the EFR is 0.5"$\times$0.5".

\section{Temporal evolution of an emerging flux loop}
In this section, we study more quantitatively the temporal evolution of individual emerging loops.

\subsection{Evolution of the footpoints}
\begin{figure}
\begin{center}
\FigureFile(80mm,120mm){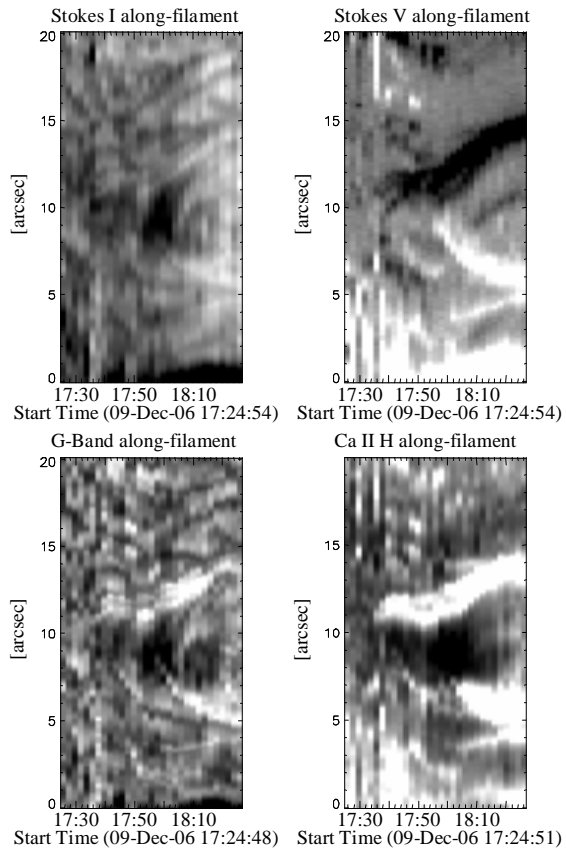}
\end{center}
\caption{
Along-filament time-sliced diagram of the emerging flux.

Upper left: Stokes I, upper right: Stokes V, lower left: G-band, lower right: Ca\emissiontype{II} H.
}\label{fig6}
\end{figure}

Fig. \ref{fig6} is the "along-filament" time-sliced diagram of an emerging loop.
Temporal variation of linear sizes and mutual separation speed of footpoints measured in Stokes I and V,
G-band and Ca\emissiontype{II} H are shown in Fig. \ref{fig8}.

\begin{figure}
\begin{center}
\FigureFile(80mm,120mm){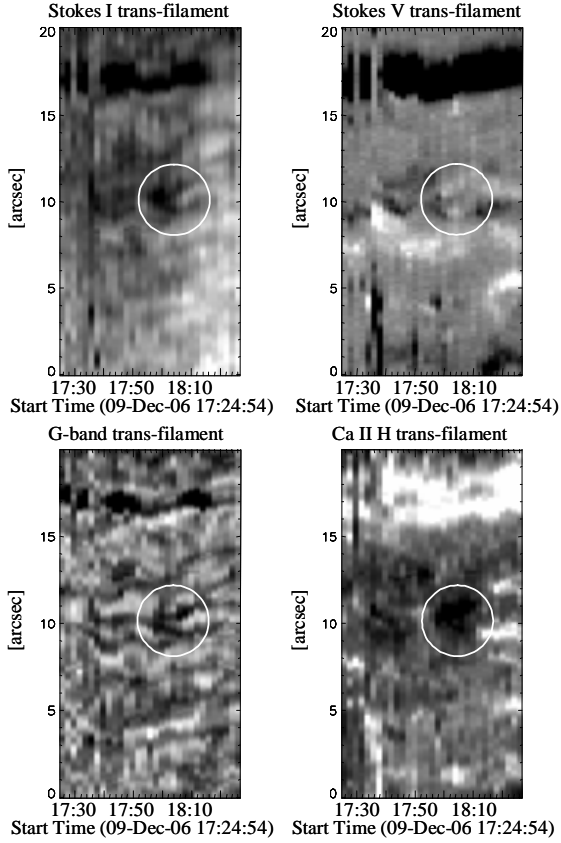}
\end{center}
\caption{
Trans-filament time-sliced diagram of the emerging flux.
Upper left: Stokes I, upper right: Stokes V, lower left: G-band, lower right: Ca\emissiontype{II} H.
Circles indicate the location of flux tube expansion.
}\label{fig7}
\end{figure}

\begin{figure}
\begin{center}
\FigureFile(80mm,60mm){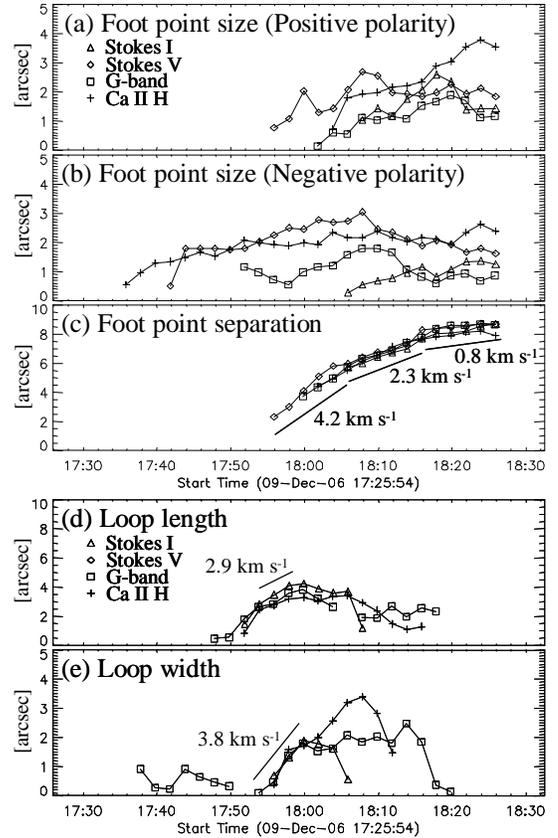}
\end{center}
\caption{
Temporal evolutions of measured quantities derived from time-sliced diagrams.
(a) Footpoint size of positive polarities.
(b) Footpoint size of negative polarities.
(c) Footpoint separation.
(d) Loop length.
(e) Loop width.
}\label{fig8}
\end{figure}

From Fig. \ref{fig8}a, we can say that Ca bright point is larger than that of Stokes I/V and G-band features. 
From Fig. \ref{fig8}b, the same result is obtained, except Stokes V feature.
As can be seen in Stokes V map of Fig. \ref{fig6}, there was a pre-existing negative polarity features on the slit.
So the size increment of negative polarity features after the emergence correspond to real size of the footpoint of emerged loop.
If so, we may conclude that the size of the Ca feature is larger than those of the photospheric features in both polarities.

The footpoints of opposite polarity move away from each other at a speed of 4.2 km s$^{-1}$ in the initial phase,
and slow down to 0.8 km s$^{-1}$ (Fig. \ref{fig8}c), which is consistent with previous work by \citet{Har73}.

\subsection{Evolution of the loops}
The along-filament time-sliced diagram of Stokes I (Fig. \ref{fig6}) clearly shows the dark lane which appeared at 17:54 UT and faded out at 18:02 UT.
So the lifetime is 8 minutes.
G-band time-sliced diagram shows the same behavior as that of the Stokes I.
The first Ca arch filament (1 in Fig. \ref{fig5}) appeared at 17:54 UT and lasted until 18:06 UT,
when in the same place two new emergence occurred (3 and 4 in Fig. \ref{fig5}).
So the lifetime of Ca arch filament 1 is 12 minutes.
The elongation speeds of Stokes I dark lane, G-band lane and Ca arch filament are of 2.9 km s$^{-1}$ (Fig. \ref{fig8}d).

From the trans-filament time-sliced diagram (Fig. \ref{fig7}), we can see the evolution of the flux width.
The dark lanes or filaments expanded laterally with the speed of 3.8 km s$^{-1}$ in all the wavelengths observed and reached the width of 2.0", although the Ca\emissiontype{II} H loop lasted longer and attained the final width of 3.5" (Fig. \ref{fig8}e).

\section{Discussion and summary}
\subsection{Discussion}
We have confirmed that the Ca bright points in the chromosphere are larger than features observed in Stokes I and G-band formed in the photosphere.
This can be expected if the decreasing gas pressure with height allows for an expansion of the flux tube with height.

\begin{figure}
\begin{center}
\FigureFile(80mm,60mm){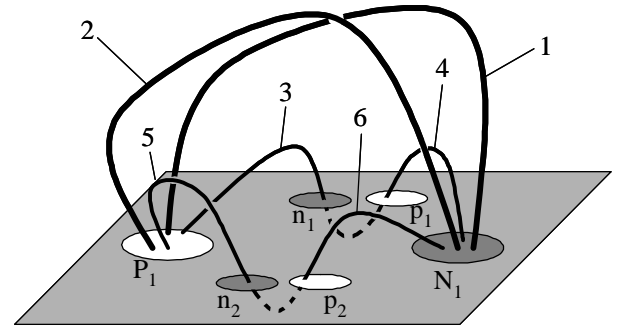}
\end{center}
\caption{
A model of the EFR.
White and dark ovals represent the footpoints observed with Stokes V.
Solid lines stand for magnetic tubes above the photosphere and dashed lines for those blow the photosphere.
The numbers and alphabets in the figure correspond to the flux tubes and the footpoints in Fig. \ref{fig5}.
}\label{fig9}
\end{figure}

Fig. \ref{fig9} shows the model of this EFR.
Temporal evolution of the observed flux emergence will be as follows:
(1) A magnetic flux tube emerges from intergranular lane (17:54 UT).
(2) The flux tube splits along its axis into two parts (flux tubes 1 and 2).
(3) After the emergence of flux tubes 1 and 2, newly emerged flux tubes appear (flux tubes 3-6).
These undulatory emergence of flux will be the ones observed by \citet{Par04} and simulated by \citet{Iso07}.

The horizontal expansion of an emerging flux tube in the photosphere was clearly detected in this work.
What is the mechanism of flux tube expansion?
This may correspond to the horizontal expansion of the emerging flux in the photosphere,
theoretically predicted by three dimensional simulations.
\citet{Mat93} simulated the emergence of non-twisted flux tube and showed that the horizontal expansion speed of flux tube is 30 km s$^{-1}$ in the photosphere.
\citet{Fan01} and \citet{Mag01} studied the emergence of twisted flux tubes and obtained the horizontal speeds of flux tube expansion to be 6-8 km s$^{-1}$.
\citet{Noz05} performed a comprehensive simulation of emerging magnetic flux sheets.
The horizontal expansion speed of non-sheared sheets is 20 km s$^{-1}$, while that of sheared sheet is $\sim$10km s$^{-1}$.
All the results of three dimensional simulations of emerging flux tubes tell us that the horizontal expansion speed will be reduced when the tubes are twisted.
As our observation showed a relatively slow speed of horizontal expansion (3.8 km s$^{-1}$) compared with the simulational results of non-twisted tubes expansion, the flux tube studied by us is probably a twisted one.

\subsection{Summary}
We observed an emerging flux which appeared near NOAA 10930 with Hinode/SOT.
From this study, the following properties about EFRs were found.

\begin{enumerate}
\item{The size of Ca bright points is larger than those of Stokes I and G-band.}
\item{Two footpoints separate each other at the speed of 4.2 km s$^{-1}$ during the initial phase of evolution.
In later phase, the separating rate decreases to 0.8 km s$^{-1}$.}
\item{The lifetime of the dark lanes in Stokes I and G-band is 8 minutes, while that of Ca filament is 12 minutes.}
\item{The Ca arch filament appears within 2 minutes after the formation of the dark lane in Stokes I image.}
\item{The width of the Ca arch filament is 3.5" at its maximum, which is wider than that of the photospheric dark lanes (2.0").}
\item{The dark lanes or filaments expand laterally with a common speed of 3.8 km $^{-1}$ in the photosphere and the chromosphere.
Slow lateral expansion of the flux tube suggests that the flux tube is a twisted one.}
\end{enumerate}

\bigskip

\section*{Acknowledgment}
The authors are partially supported by a Grant-in-Aid for the 21st Century COE 'Center for Diversity and Universality in Physics'from the Ministry of Education,
Culture, Sports, Science and Technology (MEXT) of Japan, and also partially supported by the Grant-in-Aid for 'Creative Scientific Research The Basic Study of Space
Weather Prediction' (17GS0208, Head Investigator: K. Shibata) from the Ministry of Education, Science, Sports, Technology, and Culture of Japan.
Hinode is a Japanese mission developed and launched by ISAS/JAXA, with NAOJ as domestic partner and NASA and STFC (UK) as international partners.
It is operated by these agencies in co-operation with ESA and NSC (Norway).

\end{document}